\documentclass[twocolumn,showpacs,showkeys]{revtex4}
\usepackage{graphicx}
\usepackage{bm}
\usepackage{color}
\usepackage{amsmath}
\usepackage{natbib}

\begin{document}

\title{Exchange effects in Coulomb quantum plasmas: Dispersion of waves in 2D and 3D mediums}

\author{Pavel A. Andreev}
\email{andreevpa@physics.msu.ru}
\affiliation{M. V. Lomonosov Moscow State University, Moscow, Russian Federation.}

 \date{\today}

\begin{abstract}
We describe quantum hydrodynamic equations with the Coulomb exchange interaction for three and two dimensional plasmas. Explicit form of the force densities are derived. We present non-linear Schrodinger equations (NLSEs) for the Coulomb quantum plasmas with the exchange interaction. We show contribution of the exchange interaction in the dispersion of the Langmuir, and ion-acoustic waves. We consider influence of the spin polarization ratio on strength of the Coulomb exchange interaction. This is important since exchange interaction between particles with same spin direction and particles with opposite spin directions are different. At small particle concentrations $n_{0}\ll10^{25}cm^{-3}$ and small polarization the exchange interaction gives small decrease of the Fermi pressure. With increasing of polarization role of the exchange interaction becomes more important, so that it can overcome the Fermi pressure. The exchange interaction also decreases contribution of the Langmuir frequency. Ion-acoustic waves do not exist in limit of large polarization since the exchange interaction changes the sign of pressure. At large particle concentrations $n_{0}\gg10^{25}cm^{-3}$ the Fermi pressure prevails over the exchange interaction for all polarizations. Similar picture we obtain for two dimensional quantum plasmas.
\end{abstract}


\maketitle




\section{\label{sec:level1} Introduction}

Exchange interaction is a remarkable example of quantum physical effects. The exchange interaction is related to overlapping of the wave functions of interacting particles. Hence it reveals itself as a short range interaction, even when we consider the exchange part of a long range interaction.

Let us now give a brief historical overview of quantum plasma research tracing contribution of the exchange interaction. Hydrodynamic description of quantum plasmas was considered in 1999 by Kuz'menkov and Maksimov \cite{MaksimovTMP 1999} \textit{and} Kuzelev and Rukhadze \cite{Kuzelev Ruhadze UFN1999}, where were considered spinless Coulomb quantum plasmas. The self-consistent field approach was in the center of attention of these papers. Nevertheless a general form of the exchange interaction for bosons and fermions were derived in Ref. \cite{MaksimovTMP 1999}. In 2000-2001 attention shifted towards spin-1/2 quantum plasmas \cite{Maksimov Izv 2000}-\cite{MaksimovTMP 2001 b}. The explicit form of the exchange interaction for the Coulomb and the spin-spin interactions were derived in Ref. \cite{MaksimovTMP 2001 b}. Contribution of these interactions in properties of many-electron atoms was described there.

During period 2001-2007 most of researches considered spinless
quantum plasmas \cite{Haas PRE 00}, \cite{Manfredi PRB 01} (for
review see Refs. \cite{Shukla PhUsp 2010}, \cite{Shukla RMP 11}).
A great interest to quantum plasmas of spinning particles arose
since 2007 Marklund and Brodin drew attention to the field by
their papers \cite{Marklund PRL07}, \cite{Brodin NJP 07}, some
examples we are presented below.

Dispersion properties of spin-1/2 quantum plasmas have been under
consideration for many years (see Refs. \cite{Maksimov VestnMSU
2000}, \cite{Marklund PRL07}, \cite{Harabadze RPJ 04}-\cite{Shahid PP 13}. It was shown that
magnetic moment (spin) evolution in quantum plasmas leads to new
branches of wave dispersion \cite{Vagin Izv RAN 06}, \cite{Andreev
VestnMSU 2007}, \cite{Brodin PRL 08}, \cite{Misra JPP 10},
\cite{Andreev IJMP 12}, \cite{pavelproc}, \cite{Andreev Asenjo
13}. The last of these Refs. \cite{Andreev Asenjo 13} shows
dispersion of spin wave in the spin-1/2 two dimensional electron
gas. Contribution of the exchange interaction in dispersion
properties of spin-1/2 quantum plasmas was considered in 2008 (see
Ref. \cite{Andreev AtPhys 08}). It was demonstrated that the
exchange interactions give potential force field depending on the
particle concentration and the spin density. In linear
approximation the exchange interactions can be combined with the
Fermi pressure, so we have an effective shifted Fermi velocity
\cite{Andreev AtPhys 08}. Description of spin-plasma wave
propagating perpendicular to an external magnetic field was given
in 2006 by Vagin et al. in 2006 (see Ref. \cite{Vagin Izv RAN
06}), Andreev and Kuz'menkov in 2007 (see Ref. \cite{Andreev
VestnMSU 2007}), and Brodin et al. in 2008 (see Ref. \cite{Brodin
PRL 08}). In Refs. \cite{Vagin Izv RAN 06} and \cite{Brodin PRL
08} the wave is considered in terms of kinetic equation, the
quantum hydrodynamics was applied for consideration of the wave in
Ref. \cite{Andreev VestnMSU 2007}. Spin propagating parallel to an
external magnetic field was obtained by Misra et al. in 2010 (see
Ref. \cite{Misra JPP 10}) and Andreev and Kuz'menkov in 2011 (see
Ref. \cite{pavelproc}). Influence of the spin-orbit interaction on
these waves was considered in 2011-2012 (see Refs. \cite{Andreev
IJMP 12} and \cite{pavelproc}).

Charge of particles creating the electric field is essential for
the plasma wave existence. However the spin waves can exist in
systems of neutral particles, since the magnetic field plays main
role in spin wave propagation. It was suggested in Ref.
\cite{Andreev VestnMSU 2007} that the spin-1/2 quantum plasmas can
support spin waves propagating by magnetic field with no electric
field contributing in their propagation. Dispersion of such waves
was considered in Refs. \cite{Andreev VestnMSU 2007} and
\cite{Andreev IJMP 12}. Three branches of these waves were
obtained. Dispersion of one of them depends on the Fermi velocity.
Hence the exchange interaction gives contribution in this wave
dispersion via the shifted Fermi velocity \cite{Andreev AtPhys
08}. Spin waves in systems of neutral particles existing due to
the long range spin-spin interaction are considered in Ref.
\cite{Andreev kinetics 13} in terms of quantum kinetics. This
quantum kinetics was derived in Ref. \cite{Andreev kinetics 13} as
direct generalization of the method of many-particle quantum
hydrodynamics \cite{MaksimovTMP 1999}, \cite{MaksimovTMP 2001},
\cite{MaksimovTMP 2001 b}, \cite{Andreev IJMP 12},
\cite{Trukhanova}, \cite{Andreev RPJ 07}, \cite{Andreev PRB 11}.
This kinetics differs from recent generalization of the Wigner
kinetics for spinning particles \cite{Marklund TTSP 11}-\cite{Haas NJP 10}.

All mentioned effects are based on scalar g-factor theories.
Effect of tensor g-factor on the spectrum of eigen modes in
spin-1/2 quantum plasmas was considered by Vagin et al (see Ref.
\cite{Vagin 09}).

Let us mention excellent applications of exchange interaction in
terms of many-particle quantum hydrodynamics. They were presented
in Ref. \cite{Andreev PRA08} for system of neutral quantum
particles with the short range interaction. The famous
Gross-Pitaevskii equation, for the inhomogeneous non-ideal
Bose-Einstein condensates, together with its analog for ultracold
fermions, and their generalizations were derived there.

Recent achievements in field of quantum plasmas with the exchange interactions can be found in Refs. \cite{Zamanian PRE 13 exch} and \cite{Zamanian 14 exch}, where authors applied the Wigner kinetics \cite{Wigner PR 84}.

In this paper we consider contribution of the Coulomb exchange interaction in spectrum of quantum plasma waves including ion-acoustic waves. Recently a model for classic and quantum plasmas including the finite size of ions was developed (see Ref. \cite{Andreev 1401 finite ions}). Contribution of the finite size of ions in dispersion of ion-acoustic waves was obtained in Ref. \cite{Andreev 1401 finite ions}.

This paper is organized as follows. Two fluid quantum hydrodynamics for Coulomb plasmas with the exchange interaction is presented in Sec. II. In Sec. III we give applications of the developed model to the Langmuir and ion-acoustic waves in two- and three dimensional quantum plasmas. In Sec. IV brief summary of obtained results is presented.

\section{\label{sec:level1} Model}

In this paper we present a set of quantum hydrodynamic equations
for spinless Coulomb quantum plasmas without derivation. Method of
direct derivation of the quantum hydrodynamic equations from
many-particle Schrodinger equation can be found in Refs.
\cite{MaksimovTMP 1999} (Coulomb interaction with the exchange
part for Bose and Fermi particles), \cite{MaksimovTMP 2001},
\cite{MaksimovTMP 2001 b}, \cite{Andreev RPJ 07} (spin-1/2 charged
particles), \cite{Andreev PRB 11} (charged particles baring
electric dipole moment), \cite{Ivanov Darwin} (spinless
semi-relativistic quantum plasmas). Some details of derivation can
be also found in Ref. \cite{Andreev PRA08} dedicated to quantum
neutral particles with exchange interactions.

The QHD equations for electron subsystem in electron-ion quantum plasmas
\begin{equation}\label{EXCHANGE cont eq electrons}
\partial_{t}n_{e}+\nabla(n_{e}\textbf{v}_{e})=0, \end{equation}
and
$$m_{e}n_{e}(\partial_{t}+\textbf{v}_{e}\nabla)\textbf{v}_{e}+\nabla p_{e}-\frac{\hbar^{2}}{2m_{e}}n_{e}\nabla\Biggl(\frac{\triangle n_{e}}{n_{e}}-\frac{(\nabla n_{e})^{2}}{2n_{e}^{2}}\Biggr)$$
$$=q_{e}n_{e}\biggl(\textbf{E}_{ext}+\frac{1}{c}[\textbf{v}_{e},\textbf{B}_{ext}]\biggr)-q_{e}^{2}n_{e}\nabla\int G(\textbf{r},\textbf{r}')n_{e}(\textbf{r}',t)d\textbf{r}'$$
\begin{equation}\label{EXCHANGE Euler eq electrons} -q_{e}q_{i}n_{e}\nabla\int G(\textbf{r},\textbf{r}')n_{i}(\textbf{r}',t)d\textbf{r}'+\textbf{F}_{C,e},\end{equation}
\textit{and} the QHD equations for ions
\begin{equation}\label{EXCHANGE cont eq ions}
\partial_{t}n_{i}+\nabla(n_{i}\textbf{v}_{i})=0, \end{equation}
and
$$m_{i}n_{i}(\partial_{t}+\textbf{v}_{i}\nabla)\textbf{v}_{i}+\nabla p_{i}-\frac{\hbar^{2}}{2m_{i}}n_{i}\nabla\Biggl(\frac{\triangle n_{i}}{n_{i}}-\frac{(\nabla n_{i})^{2}}{2n_{i}^{2}}\Biggr)$$
$$=q_{i}n_{i}\biggl(\textbf{E}_{ext}+\frac{1}{c}[\textbf{v}_{i},\textbf{B}_{ext}]\biggr)-q_{i}^{2}n_{i}\nabla\int G(\textbf{r},\textbf{r}')n_{i}(\textbf{r}',t)d\textbf{r}'$$
\begin{equation}\label{EXCHANGE Euler eq ions} -q_{e}q_{i}n_{i}\nabla\int G(\textbf{r},\textbf{r}')n_{e}(\textbf{r}',t)d\textbf{r}'+\textbf{F}_{C,i}.\end{equation}
Set of equations (\ref{EXCHANGE cont eq electrons}), (\ref{EXCHANGE Euler eq electrons}) and (\ref{EXCHANGE cont eq ions}), (\ref{EXCHANGE Euler eq ions}) are coupled to each other by means of the last terms in the Euler equations (\ref{EXCHANGE Euler eq electrons}) and (\ref{EXCHANGE Euler eq ions}). In equation set (\ref{EXCHANGE cont eq electrons})-(\ref{EXCHANGE Euler eq ions}) we assumed that thermal pressure is isotropic:
$p_{a}^{\alpha\beta}=p_{a}\delta^{\alpha\beta}$, where $a$ stands for species of particles. Equations (\ref{EXCHANGE cont eq electrons}) and (\ref{EXCHANGE cont eq ions}) are continuity equations for electrons and ions correspondingly. These equations show conservation of particle number of electrons and ions. Equations (\ref{EXCHANGE Euler eq electrons}) and (\ref{EXCHANGE Euler eq ions}) are the momentum balance (Euler) equations for electrons and ions. The first terms in the left-hand side of Euler equations $m_{a}n_{a}(\partial_{t}+\textbf{v}_{a}\nabla)\textbf{v}_{a}$ are the kinematic part. The second terms are the gradient of the thermal pressure or the Fermi pressure for degenerate electrons and ions. It appears as the thermal part of the momentum flux related to distribution of particles on states with different momentum. The third terms are the quantum Bohm potential appearing as the quantum part of the momentum flux. In the right-hand sides of the Euler equations we present interparticle interaction and interaction of particles with external electromagnetic fields. The first group of terms in the right-hand side of the Euler equations describe interaction with the external electromagnetic fields. The second term in the Euler equation for electrons (\ref{EXCHANGE Euler eq electrons}) describes the electron-electron Coulomb interaction. The third term is the Coulomb action of ions on electron motion. The second term in the Euler equation for ions (\ref{EXCHANGE Euler eq ions}) gives the ion-ion Coulomb interaction. The Coulomb field of electrons acting on ions is presented by the third term in equation (\ref{EXCHANGE Euler eq ions}). The last terms in the right-hand sides of equations (\ref{EXCHANGE Euler eq electrons}) and (\ref{EXCHANGE Euler eq ions}) describes the Coulomb electron-electron and ion-ion exchange interactions $\textbf{F}_{C,a}$ correspondingly. These two terms are main subject of this paper. Below we consider their contribution in the plasma wave dispersion.

We can rewrite  the Euler equations (\ref{EXCHANGE Euler eq electrons}) and (\ref{EXCHANGE Euler eq ions}) in terms of the self-consistent electric field
\begin{equation}\label{EXCHANGE } \textbf{E}_{a(int)}=-q_{a}\nabla\int G(\textbf{r},\textbf{r}')n_{a}(\textbf{r}',t)d\textbf{r}',\end{equation}
where $\textbf{E}_{a(int)}$ is the electric field created by particles of species $a=e,i$. Hence equations (\ref{EXCHANGE Euler eq electrons}), (\ref{EXCHANGE Euler eq ions}) attain more familiar form
$$m_{e}n_{e}(\partial_{t}+\textbf{v}_{e}\nabla)\textbf{v}_{e}+\nabla p_{e}-\frac{\hbar^{2}}{2m_{e}}n_{e}\nabla\Biggl(\frac{\triangle n_{e}}{n_{e}}-\frac{(\nabla n_{e})^{2}}{2n_{e}^{2}}\Biggr)$$
\begin{equation}\label{EXCHANGE Euler eq electrons non INT} =q_{e}n_{e}\biggl(\textbf{E}_{ext}+\textbf{E}_{int}+\frac{1}{c}[\textbf{v}_{e},\textbf{B}_{ext}]\biggr)+\textbf{F}_{C,e},\end{equation}
\textit{and} the Euler equation for ions
$$m_{i}n_{i}(\partial_{t}+\textbf{v}_{i}\nabla)\textbf{v}_{i}+\nabla p_{i}-\frac{\hbar^{2}}{2m_{i}}n_{i}\nabla\Biggl(\frac{\triangle n_{i}}{n_{i}}-\frac{(\nabla n_{i})^{2}}{2n_{i}^{2}}\Biggr)$$
\begin{equation}\label{EXCHANGE Euler eq ions non INT} =q_{i}n_{i}\biggl(\textbf{E}_{ext}+\textbf{E}_{int}+\frac{1}{c}[\textbf{v}_{i},\textbf{B}_{ext}]\biggr)+\textbf{F}_{C,i}.\end{equation}
Physical meaning of terms in the non-integral Euler equations (\ref{EXCHANGE Euler eq electrons non INT}) and (\ref{EXCHANGE Euler eq ions non INT}) is similar to described above for equations (\ref{EXCHANGE Euler eq electrons}) and (\ref{EXCHANGE Euler eq ions}). Interparticle interactions  are presented in (\ref{EXCHANGE Euler eq electrons non INT}) and (\ref{EXCHANGE Euler eq ions non INT}) in terms of internal electric field satisfying the Maxwell equations.

The Maxwell equations $\nabla\times \textbf{E}_{int}=0$ and $\nabla \textbf{E}_{int}=4\pi\rho$, where $\rho_{3D}=\sum_{a}q_{a}n_{a}$, and $\rho_{2D}=\delta(z)\sum_{a}q_{a}n_{a, (2D)}$, with $\textbf{E}_{int}=\sum_{a}\textbf{E}_{a(int)}$. Presented here two dimensional charge density $\rho_{2D}=\rho_{2D}(x,y,z)$ is explicitly presented as two dimensional layer in the three dimensional physical space. Two dimensional particle concentration is a function of two space coordinates $n_{a, (2D)}=n_{a, (2D)}(x,y)$ in plane $z=0$.

Equation of state for degenerate 2D Fermi gas is
$p_{a,2D}=\pi\hbar^{2}n_{a,2D}^{2}/(2m_{a})$. In 3D case one
similarly finds
$p_{a,3D}=(3\pi^{2})^{2/3}\hbar^{2}n_{a,3D}^{5/3}/(5m_{a})$. These
equations of state take place for unpolarized fermions at zero
temperature $p_{a,ND}=p_{a,ND\uparrow\downarrow}$, $N=2$ or $3$,
where subindex $\uparrow\downarrow$ means that in each occupied
quantum state we have two particles with opposite spins. Hence we
have two fermions in each state with energy lower than the Fermi
energy $\varepsilon_{F}$, fermions of each pair have opposite
spins. When system of spin-1/2 fermions is polarized then
distribution of fermions looks like one electron in each state
with energy lower than $2^{2/3}\varepsilon_{F,3D}$ for 3D mediums,
and $2\varepsilon_{F,2D}$ for 2D mediums. For polarized systems
equations of state appears as
$p_{a,3D\uparrow\uparrow}=2^{2/3}(3\pi^{2})^{2/3}\hbar^{2}n_{a,3D}^{5/3}/(5m_{a})$
for 3D mediums, and
$p_{a,2D\uparrow\uparrow}=2\pi\hbar^{2}n_{a,2D}^{2}/(2m_{a})$ for
2D mediums, where subindex $\uparrow\uparrow$ means that all
particles have same spin direction. We may consider partly
polarized particles, then we need to introduce ratio of
polarizability $\eta=\frac{\mid
n_{\uparrow}-n_{\downarrow}\mid}{n_{\uparrow}+n_{\downarrow}}$,
with indexes $\uparrow$ and $\downarrow$ means particles with spin
up and spin down. Here we have that instead
$p_{0}=p_{\uparrow}+p_{\downarrow}$, with
$p_{\uparrow}=p_{\downarrow}=p_{0}/2$ for $\eta=0$, we find
$p=\tilde{p}_{\uparrow}+\tilde{p}_{\downarrow}$, with
$\tilde{p}_{\uparrow}=2^{5/3}p_{\uparrow}=2^{2/3}p_{0}$ and
$\tilde{p}_{\downarrow}=0$ for $\eta=1$ in 3D case. Similarly we
have $p_{0}=p_{\uparrow}+p_{\downarrow}$, with
$p_{\uparrow}=p_{\downarrow}=p_{0}/2$ at $\eta=0$, we obtain for
$\eta=1$ $p=\tilde{p}_{\uparrow}+\tilde{p}_{\downarrow}$, with
$\tilde{p}_{\uparrow}=4p_{\uparrow}=2p_{0}$ and
$\tilde{p}_{\downarrow}=0$ for 2D case. In general case of
partially polarized system of particles we can write
$p_{a,3D\Updownarrow}=\vartheta_{3D}(3\pi^{2})^{2/3}\hbar^{2}n_{a,3D}^{5/3}/(5m_{a})$
for 3D mediums, and
$p_{a,2D\Updownarrow}=\vartheta_{2D}\pi\hbar^{2}n_{a,2D}^{2}/(2m_{a})$
for 2D mediums, with
\begin{equation}\label{EXCHANGE} \vartheta_{3D}=\frac{1}{2}[(1+\eta)^{5/3}+(1-\eta)^{5/3}],\end{equation}
and
\begin{equation}\label{EXCHANGE} \vartheta_{2D}=1+\eta^{2},\end{equation}
where $\Updownarrow$ stands for partially polarized systems, that
means that part of states contain two particle with opposite spins
and other occupied states contain one particle with same spin
direction.

Considering two electrons one finds that full wave function is anti-symmetric. If one has two electrons with parallel spins one has that wave function is symmetric on spin variables, so it should be anti-symmetric on space variables. In opposite case of anti-parallel spins one has anti-symmetry of wave function on spin variables and symmetry of wave function on space variables.

Considering energy of two electron Coulomb interaction one finds it two parts: the classic like part C and the exchange part A. For the parallel (anti-parallel) spins one obtains E$_{\uparrow\uparrow}$=C-A (E$_{\uparrow\downarrow}$=C+A). Parallel (anti-parallel) configuration of spins decreases (increases) energy of the Coulomb interaction.

Systems of unpolarized electrons then average numbers of electrons with different direction of spins equal to each other, we find that average number of particles for a chosen with parallel and anti-parallel spins is the same. Consequently we have that average exchange interaction equals to zero.

In partly polarized systems the numbers of particles with different spin are not the same. In this case a contribution of the average exchange interaction appears. At full measure it reveals in fully polarized system then all electrons have same direction of spins. In accordance with the previous discussion we find that exchange interaction, for this configuration, gives attractive contribution in the force field.

This result allows us to find the force field of the Coulomb exchange interaction
$$\textbf{F}_{C,a(3D)}=2^{4/3} q_{a}^{2}\sqrt[3]{\frac{3}{\pi}}\sqrt[3]{n_{a}}\nabla n_{a}$$
\begin{equation}\label{EXCHANGE F C exchange 3D} = 6q_{a}^{2}\sqrt[3]{\frac{6}{\pi}}  n_{a} \nabla n_{a}^{1/3}. \end{equation}

We obtain the force field of exchange interaction for 2D quantum plasmas in the following form
$$\textbf{F}_{C,a(2D)}=2^{3/2} \sqrt{2\pi}\frac{24 \textrm{arsh}1}{\pi^{2}}q_{a}^{2}\sqrt{n_{a}}\nabla n_{a}$$
\begin{equation}\label{EXCHANGE F C exchange 2D} =  8\sqrt{\pi}\frac{\beta}{\pi^{2}}q_{a}^{2} n_{a}\nabla \sqrt{n_{a}}.\end{equation}
where we introduce $\beta\equiv 24 \textrm{arsh}1=21.153$.

The force fields (\ref{EXCHANGE F C exchange 3D}) and (\ref{EXCHANGE F C exchange 2D}) are obtained for fully polarized systems of identical particles.

For partially polarized particles the force fields reappear as $\textbf{F}_{C,a(3D)}=\zeta_{3D} q_{a}^{2}\sqrt[3]{\frac{3}{\pi}}\sqrt[3]{n_{a}}\nabla n_{a}$ and $\textbf{F}_{C,a(2D)}=\zeta_{2D}  \sqrt{2\pi}\frac{\beta}{\pi^{2}}q_{a}^{2}\sqrt{n_{a}}\nabla n_{a}$, with
\begin{equation}\label{EXCHANGE} \zeta_{3D}=(1+\eta)^{4/3}-(1-\eta)^{4/3}\end{equation}
and
\begin{equation}\label{EXCHANGE} \zeta_{2D}=(1+\eta)^{3/2}-(1-\eta)^{3/2}.\end{equation}
We should mention that coefficients $\zeta_{3D}\sim\eta$ and $\zeta_{2D}\sim\eta$ are proportional to spin polarization. Limit cases of $\zeta_{3D}$ and $\zeta_{2D}$ are $\zeta_{3D}(0)=0$, $\zeta_{3D}(1)=2^{4/3}$, $\zeta_{2D}(0)=0$ and $\zeta_{2D}(1)=2^{3/2}$.

If we do not apply a magnetic field to systems under consideration we need to have systems, where self-organization of equilibrium state leads to formation of population levels with uncompensated spin, such it is in ferromagnetic domains. Overwise we have no contribution of the Coulomb exchange interaction.

Force fields of exchange interaction (\ref{EXCHANGE F C exchange 3D}) and (\ref{EXCHANGE F C exchange 2D}) are potential fields. Thus they do not give contribution in dispersion of transverse waves. They affect longitudinal waves and waves with complex polarization: longitudinal-transverse waves.
Consequently electromagnetic waves are not affected by the Coulomb exchange interaction in 3D and 2D mediums in absence of external fields.

Force fields of the Coulomb exchange interaction can be presented as product of the particle concentration on the gradient of a function when it gives no contribution in equations of the vorticity evolution (see Refs. \cite{Andreev Asenjo 13}, \cite{Mahajan PRL 11}, \cite{Braun PRL 12}, and \cite{Mahajan PL A 13}).

Many-particle quantum hydrodynamic equations can be represented in form of the non-linear Schrodinger equation (NLSE). Let us consider evolution of electrons at motionless ions. We introduce the wave function in medium defined in terms of hydrodynamic variables
\begin{equation}\label{EXCHANGE wave funct in medium} \Phi=\sqrt{n}\exp\biggl(\imath \frac{mS}{\hbar}\biggr),\end{equation}
where $S$ is the potential of velocity field. Let us mention that the NLSE can be derived for eddy-free motion of electrons. Definition (\ref{EXCHANGE wave funct in medium}) can be applied for three dimensional and low dimensional systems of particles $\mid\Phi_{3D}\mid=\sqrt{n}_{3D}$, and $\mid\Phi_{2D}\mid=\sqrt{n}_{2D}$, with $[n_{3D}]=cm^{-3}$, $[n_{2D}]=cm^{-2}$. To derive NLSE we need to differentiate function (\ref{EXCHANGE wave funct in medium}) with respect to time. After some calculations we find NLSE for electrons in three dimensional quantum plasmas
$$\imath\hbar\partial_{t}\Phi_{3D}=\biggl(-\frac{\hbar^{2}\triangle}{2m_{e}}+\vartheta_{3D}\frac{(3\pi^{2})^{2/3}\hbar^{2}n_{e}^{2/3}}{2m_{e}}$$
\begin{equation}\label{EXCHANGE NLSE 3D}
-3\sqrt[3]{\frac{3}{\pi}}\zeta_{3D}q^{2}n^{1/3}+q_{e}\varphi\biggr)\Phi_{3D}.\end{equation}
For 2DEGs we obtain
$$\imath\hbar\partial_{t}\Phi_{2D}=\biggl(-\frac{\hbar^{2}\triangle}{2m_{e}}+ \vartheta_{2D}\frac{\pi\hbar^{2}}{m_{e}}n$$
\begin{equation}\label{EXCHANGE NLSE 2D}
-\frac{2\beta \sqrt{2\pi}}{\pi^{2}}\zeta_{2D}q_{e}^{2}\sqrt{n}+q_{e}\varphi\biggr)\Phi_{2D}.\end{equation}
Equations (\ref{EXCHANGE NLSE 3D}) and (\ref{EXCHANGE NLSE 2D}) contain potential of the electric field $\varphi$ presenting the external and internal electric fields: $\textbf{E}=-\nabla\varphi$.

Considering dynamic of two or more species we should present the wave functions in medium for each species and derive NLSEs for each species either.

\section{\label{sec:level1} Applications}

In this section we consider small perturbations of equilibrium state describing by nonzero particle concentration $n_{0}$, and zero velocity field $\textbf{v}_{0}=0$ and electric field $\textbf{E}_{0}=0$.

Assuming that perturbations are monochromatic
\begin{equation}\label{EXCHANGE perturbations}
\left(\begin{array}{ccc} \delta n
 \\
\delta \textbf{v}
 \\
\delta \textbf{E}
\\
\end{array}\right)=
\left(\begin{array}{ccc}
N_{A} \\
\textbf{V}_{A} \\
\textbf{E}_{A}\\
\end{array}\right)e^{-\imath\omega t+\imath \textbf{k} \textbf{r}},\end{equation}
we get a set of linear algebraic equations relatively to $N_{A}$ and $V_{A}$. Condition of existence of nonzero solutions for amplitudes of perturbations gives us a dispersion equation.

\subsection{Three dimensional quantum plasmas}

In classic plasmas the Langmuir waves have the following spectrum
\begin{equation}\label{EXCHANGE spectrum Langm 3D} \omega^{2}=\omega_{Le}^2+\frac{\gamma T}{m_{e}}k^2, \end{equation}
with the three dimensional Langmuir frequency
\begin{equation}\label{EXCHANGE Langmuir freq 3D} \omega_{Le,3D}^{2}=\frac{4\pi e^2 n_{0,3D}}{m_{e}},\end{equation}
$T$ is the temperature, $\gamma$ is the adiabatic index.

We consider quantum plasmas, so we are interested in the low temperature properties, then carriers are degenerated. Hence we have contribution of the Fermi pressure instead of the temperature.

Spectrum of the Langmuir waves in 3D quantum plasmas with the Coulomb exchange interaction appears as
$$\omega^{2}=\omega_{Le, 3D}^{2}-\zeta_{3D}\sqrt[3]{\frac{3}{\pi}}\frac{e^{2}}{m_{e}}\sqrt[3]{n_{0e}}k^{2}$$
\begin{equation}\label{EXCHANGE disp Lang 3D} +\vartheta_{3D}\frac{(3\pi^{2})^{2/3}\hbar^{2}n_{0e}^{2/3}}{3m_{e}^{2}}k^{2}+\frac{\hbar^{2}k^{4}}{4m_{e}^{2}}. \end{equation}
The first term in formula (\ref{EXCHANGE disp Lang 3D}) is the 3D
Langmuir frequency existing due to the Coulomb interaction in the
self-consistent field approximation. The second term describes the
Coulomb exchange interaction between electrons. The third term
appears from the Fermi pressure. The last term describes
contribution of the quantum Bohm potential.

Let us consider a dimensionless form of the spectrum of 3D Langmuir waves. Introducing dimensionless frequency $\Omega=\omega/\omega_{Le,3D}$ and wave vector $\xi=k/\sqrt[3]{n_{0e,3D}}$ we obtain
$$\Omega^{2}=1-\zeta_{3D}\frac{1}{4\pi}\sqrt[3]{\frac{3}{\pi}}\xi^{2}$$
\begin{equation}\label{EXCHANGE disp Lang 3D dim-less} +\vartheta_{3D}\biggl(\frac{(3\pi^{2})^{2/3}}{3}+\frac{1}{16\pi}\xi^{2}\biggr)\Lambda_{3D}\xi^{2}, \end{equation}
where we also have a dimensionless parameter
\begin{equation}\label{EXCHANGE disp Lang 3D a dim-less parametr} \Lambda_{3D}=\frac{\hbar^{2}}{m_{e}e^{2}}\sqrt[3]{n_{0e}}, \end{equation}
depending on fundamental constants $\hbar$, $e$, $m_{e}$, and the equilibrium concentration of electrons $\sqrt[3]{n_{0e,3D}}$. The contribution of the exchange interaction can be considered as a shift of the Fermi pressure since both of them are proportional to $\xi^{2}$. On the other hand, the term describing the exchange interaction is proportional to the square of the Langmuir frequency. From this point of view we see that exchange interaction gives a shift of the Langmuir frequency and this shift is proportional to square of the wave vector $\xi^{2}$. So exchange interaction is considerable in the short wave length limit.

Increasing of the particle concentration allows to increase maximal wave vector of wave propagation in the medium. Hence, in short wave length limit the first two terms reveal same dependence on the equilibrium particle concentration. For whole range of wave vectors we see that the first term $\sim n_{0e,3D}$ grows faster than the second term $\sim n_{0e,3D}^{1/3}$ with the increasing of the particle concentration. The third term has an intermediate rate of grow being proportional  to $n_{0e,3D}^{2/3}$. However, in the short wave length limit the third and fourth terms increase rather faster like $n_{0e,3D}^{2/3}k^{2}$ and $k^{4}$ correspondingly. Finally formula (\ref{EXCHANGE disp Lang 3D dim-less}) shows that the third and fourth terms can be rather large at high densities (particle concentrations) and large wave vectors.

Let us estimate contribution of the exchange interaction in compare with other terms in dispersion dependence of the Langmuir waves. Comparing exchange interaction with the Fermi pressure we should consider ratio of the second term to the third term
\begin{equation}\label{EXCHANGE}\chi_{EF}=\frac{m_{e}e^{2}}{\hbar^{2}\sqrt[3]{n_{0e,3D}}}=\frac{1}{\Lambda_{3D}}\approx\sqrt[3]{\frac{10^{25}}{n_{0e,3D}}}.\end{equation}
Exchange interaction prevails over the Fermi pressure than $\chi_{EF}>1$, that corresponds to $n_{0e,3D}<10^{25}$cm$^{-3}$. Hence the Coulomb exchange interaction is larger than the Fermi pressure in metals $n_{0e,3D}\sim10^{22}$$cm^{-3}$ and semiconductors $n_{0e,3D}\sim10^{18}$$cm^{-3}$. Considering extreme astrophysical objects like white dwarfs $n_{0e,3D}\sim10^{28}$$cm^{-3}$ we find that Fermi pressure is larger than the Coulomb exchange interaction. Ratio between the Fermi pressure and the exchange interaction does not depend on the wave vector $k$.

We have considered high frequency waves. Nest step is consideration of the low frequency excitations
$$\omega_{3D}(k)=kv_{s,3D}\sqrt{1-\frac{\zeta_{3D}}{\vartheta_{3D}}\frac{3}{4\pi}\sqrt[3]{\frac{3}{\pi}}\frac{\omega_{Le,3D}^{2}}{n_{0e,3D}^{2/3}v_{Fe,3D}^{2}}}\times$$
\begin{equation}\label{EXCHANGE disp ion-acoustic 3D}   \times\frac{1}{\sqrt{1+(kr_{De,3D})^{2} \biggl(1-\frac{\zeta_{3D}}{\vartheta_{3D}}\frac{3}{4\pi}\sqrt[3]{\frac{3}{\pi}}\frac{\omega_{Le,3D}^{2}}{n_{0e,3D}^{2/3}v_{Fe,3D}^{2}}\biggr)}},\end{equation}
where $v_{s,3D}=\sqrt{m_{e}/m_{i}}\sqrt{\vartheta_{3D}}\cdot v_{Fe,3D}/3$ is the three dimensional velocity of sound, $r_{De,3D}=\sqrt{\vartheta_{3D}}v_{Fe,3D}/(3\omega_{Le,3D})$ is the Debye radius. In formula (\ref{EXCHANGE disp ion-acoustic 3D}) and similar formulas below we extract contribution of the Fermi pressure. Hence formulas for ion-acoustic waves contains well-known contribution of the pressure multiplied by factor showing contribution of exchange interaction.

In the long wavelength limit we have
\begin{equation}\label{EXCHANGE IAW 3D LWL} \omega(k)=kv_{s,3D}\sqrt{1-\frac{\zeta_{3D}}{\vartheta_{3D}}\frac{3}{4\pi}\sqrt[3]{\frac{3}{\pi}}\frac{\omega_{Le,3D}^{2}}{n_{0e,3D}^{2/3}v_{Fe,3D}^{2}}}.\end{equation}

In the short wavelength limit we find
\begin{equation}\label{EXCHANGE IAW 3D SWL} \omega^{2}(k)=\omega_{Li,3D}^{2}. \end{equation}

From formula (\ref{EXCHANGE IAW 3D SWL}) we see that the exchange interaction gives no contribution in the ion-acoustic waves in the short wave length limit.

\subsection{Two dimensional quantum plasmas: two dimensional electron gas and ion motion contribution}

Two dimensional quantum plasmas are systems of electrons and ions being under confinement, so we have plane-like objects. 2D quantum plasmas are surrounded by medium. However, main properties can be obtained considering a 2D layer in empty 3D space. Such objects as 2DEG (two dimensional electron gas) and 2DHG (two dimensional hole gas) are common objects in physics of semiconductors. As application these objects appears as parts of transistors. Consideration of 2DEG and 2DHG corresponds to description of high frequency excitations. In this section we also include ion dynamics.

Spectrum of high frequency longitudinal excitations, which are the Langmuir waves, appears as follows
$$\omega^{2}=\omega_{Le, 2D}^{2}-\zeta_{2D}\frac{\beta\sqrt{2\pi}e^{2}}{\pi^{2}m_{e}}\sqrt{n_{0e}}k^{2}$$
\begin{equation}\label{EXCHANGE disp Lang 2D} +\vartheta_{2D}\frac{\pi\hbar^{2}n_{0e}}{m_{e}^{2}}k^{2}+\frac{\hbar^{2}k^{4}}{4m_{e}^{2}}, \end{equation}
where we have used the two dimensional Langmuir frequency
\begin{equation}\label{EXCHANGE Langmuir frq 2D}\omega_{Le,2D}^{2}=\frac{2\pi e^2 k n_{0,2D}}{m_{e}}\sim k,\end{equation}
which is not a constant, but it is proportional to the wave vector $k$.

Similarly to the three dimensional spectrum (see formula
(\ref{EXCHANGE disp Lang 3D})) different terms in formula
(\ref{EXCHANGE disp Lang 2D}) have the following meaning:
self-consistent Coulomb interaction, the exchange Coulomb
interaction, the Fermi pressure, and the quantum Bohm potential.

Next we discuss some properties of the 2D Langmuir wave spectrum.

Let us consider the exchange interaction with the Fermi pressure for 2D quantum plasmas. To this end we introduce the following dimensionless parameter
\begin{equation}\label{EXCHANGE} \chi_{EF,2D}=\frac{m_{e}e^{2}}{\hbar^{2}\sqrt{n_{0e,2D}}}\approx\sqrt{\frac{10^{16}}{n_{0e,2D}}}.\end{equation}
In 2D semiconductor objects $n_{0e,2D}\ll10^{16}$cm$^{-2}$. Consequently, the exchange interaction plays significant role in collective properties of semiconductors.

To this end we present a dimensionless form of formula (\ref{EXCHANGE disp Lang 2D}) introducing dimensionless parameters $\Omega_{2D}=\omega\sqrt{m/(2\pi e^{2}n_{0e,2d}^{3/2})}$ and $\xi_{2D}=k/\sqrt{n_{0e,2D}}\sim ak$, with $a$ is the average interparticle distance. Hence we have
$$\Omega_{2D}^{2}=\xi\biggl(1-\zeta_{2D}\frac{\beta}{\sqrt{2\pi}\pi^{2}}\xi\biggr)$$
\begin{equation}\label{EXCHANGE disp Lang 2D dim less} +\frac{1}{2}\vartheta_{2D}\Lambda_{2D}\xi^{2}\biggl(1+\frac{1}{4\pi}\xi^{2}\biggr), \end{equation}
with
\begin{equation}\label{EXCHANGE disp Lang 2D a dim less parametr} \Lambda_{2D}=\frac{\hbar^{2}}{me^{2}}\sqrt{n_{0e,2D}}. \end{equation}

Spectrum of the 2D ion-acoustic waves in presence of the exchange interaction appears as
$$\omega_{2D}(k)=kv_{s,2D}\sqrt{1-\frac{\zeta_{2D}}{\vartheta_{2D}}\frac{2\beta\sqrt{2\pi}e^{2}\sqrt{n_{0e,2D}}}{\pi^{2}mv_{Fe,2D}^{2}}}\times$$
\begin{equation}\label{EXCHANGE disp ion-acoustic 2D a} \times\frac{1}{\sqrt{1+(kr_{De,2D}(k))^{2}\biggl(1-\frac{\zeta_{2D}}{\vartheta_{2D}}\frac{2\beta\sqrt{2\pi}e^{2}\sqrt{n_{0e,2D}}}{\pi^{2}mv_{Fe,2D}^{2}}\biggr)}},\end{equation}
with $v_{s,2D}=\sqrt{m_{e}/m_{i}}\sqrt{\vartheta_{2D}}\cdot v_{Fe,2D}/2$ is the two dimensional velocity of sound, $r_{De,2D}=\sqrt{\vartheta_{2D}}v_{Fe,2D}/(2\omega_{Le,2D})$.
The spectrum of the 2D ion-acoustic waves can be written in more explicit form, which shows dependence on the wave vector $k$
$$\omega_{2D}(k)=kv_{s,2D}\sqrt{1-\frac{\zeta_{2D}}{\vartheta_{2D}}\frac{2\beta\sqrt{2\pi}e^{2}\sqrt{n_{0e,2D}}}{\pi^{2}mv_{Fe,2D}^{2}}}\times$$
\begin{equation}\label{EXCHANGE disp ion-acoustic 2D}   \times\frac{1}{\sqrt{1+kD\biggl(1-\frac{\zeta_{2D}}{\vartheta_{2D}}\frac{2\beta\sqrt{2\pi}e^{2}\sqrt{n_{0e,2D}}}{\pi^{2}mv_{Fe,2D}^{2}}\biggr)}},\end{equation}
where
\begin{equation}\label{EXCHANGE} D=\frac{\hbar}{2\sqrt{2\pi}e^{2}\sqrt{n_{0e,2D}}}.\end{equation}

In the long wavelength limit we have
\begin{equation}\label{EXCHANGE} \omega(k)=kv_{s,2D}\sqrt{1-\frac{\zeta_{2D}}{\vartheta_{2D}}\frac{2\beta\sqrt{2\pi}e^{2}\sqrt{n_{0e,2D}}}{\pi^{2}mv_{Fe,2D}^{2}}}.\end{equation}

In the short wavelength limit we find
\begin{equation}\label{EXCHANGE} \omega^{2}(k)=\omega_{Li,2D}^{2}\sim k. \end{equation}
As in 3D case we find no contribution of the exchange interaction in the short wave length ion-acoustic waves.

\section{Conclusions}

We have briefly described quantum hydrodynamic model for electron-ion quantum plasmas with exchange interaction. We have considered three and two dimensional electron-ion quantum plasmas. We have derived explicit form of the Coulomb exchange interaction force field for electron-electron and ion-ion Coulomb interaction. We have applied this model to dispersion properties of electron-ion plasmas tracing contribution of the exchange interaction. We have considered small amplitude linear excitations in absence of external field. However obtained results open possibilities for consideration of nonlinear effects. Derived force fields for the Coulomb exchange interaction can be used to study various effects in magnetized plasmas as well.


\begin{acknowledgements}
The author thanks Professor L. S. Kuz'menkov for fruitful discussions.
\end{acknowledgements}

\end{document}